\documentclass[
    ,final            
  ]
  {myaipproc}

\layoutstyle{6x9}

\begin{document}

\title{On the Dynamical Generation of Quark-Level-Linear-Sigma-Model-like Theories beyond one Loop}


\author{Frieder Kleefeld \footnote{e-mail: kleefeld@cfif.ist.utl.pt, URL: http://cfif.ist.utl.pt/$\sim$kleefeld/ \newline \mbox{}$\;\;$ Present address for postal correspondence: Pfisterstr.\ 31, 90762 F\"urth, Germany}}{
  address={Collaborator of the\\ Centro de F\'{\i}sica das Interac\c{c}\~{o}es Fundamentais (CFIF), Instituto Superior T\'{e}cnico,\\
Edif\'{\i}cio Ci\^{e}ncia, Piso 3, Av. Rovisco Pais, P-1049-001 LISBOA, Portugal}
} 

\begin{abstract}
A self-consistent strategy to complete the dynamical generation of Quark-Level-Linear-Sigma-Model-like Lagrangean theories beyond one loop as proposed in more detail in our manuscript arXiv:0802.1540 [hep-ph] \cite{Kleefeld:2008pn} is shortly outlined.
\end{abstract}

\maketitle

\begin{figure}
  \includegraphics[width=1.00\textwidth,height=0.08\textheight]{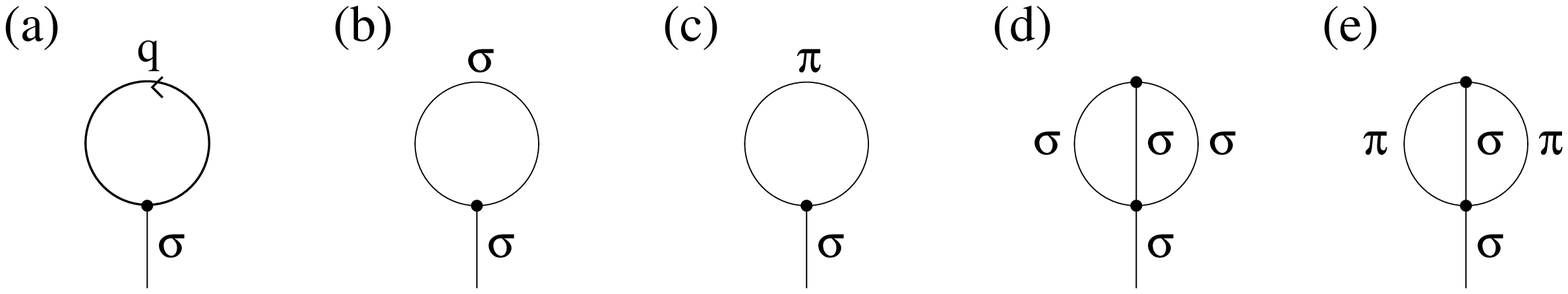}
  \caption{Tadpole sum: contributions to the $\sigma$ one-point function} \label{fig1}
\end{figure}

\begin{figure}
  \includegraphics[width=1.00\textwidth,height=0.16\textheight]{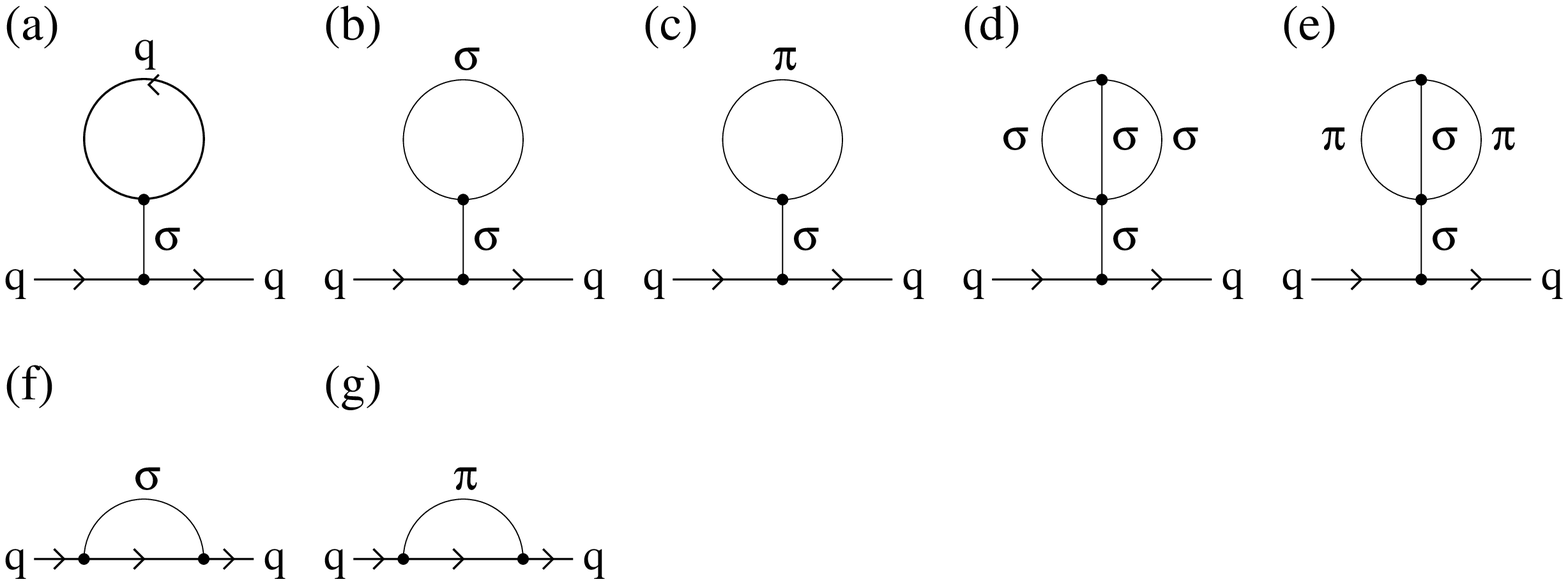}
  \caption{Quark mass: contributions to the quark self-energy} \label{fig2}
\end{figure}

\begin{figure}
  \includegraphics[width=1.00\textwidth,height=0.24\textheight]{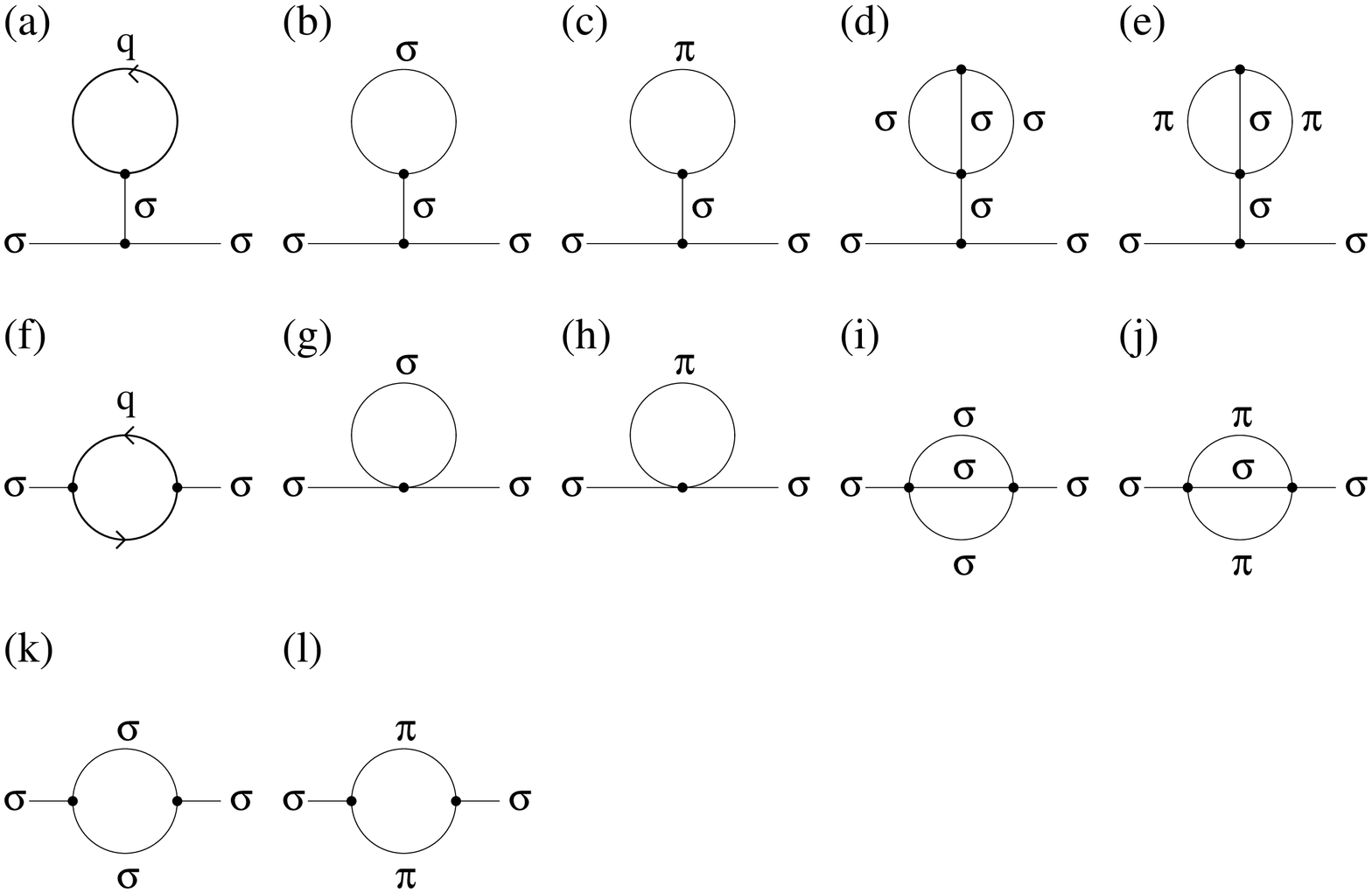}
  \caption{Sigma mass: contributions to the $\sigma$ self-energy} \label{fig3}
\end{figure}

\begin{figure}
  \includegraphics[width=1.00\textwidth,height=0.24\textheight]{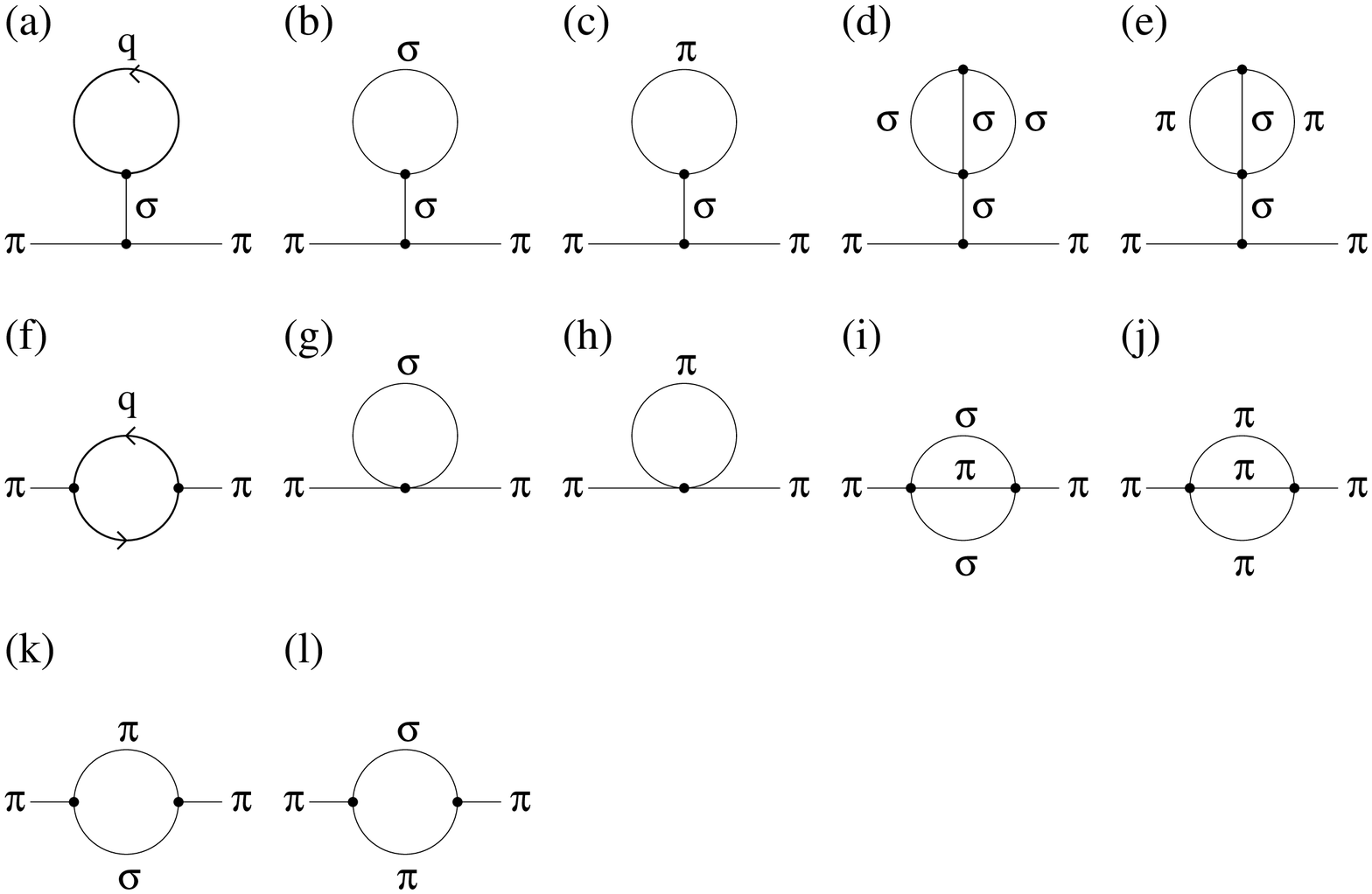}
  \caption{Pion mass: contributions to the $\pi$ self-energy} \label{fig4}
\end{figure}

\noindent Throughout the construction of \mbox{Lagrangean} densities used e.g.\ in particle physics one faces at least two problems:  \mbox{1) unpredictive} Lagrangeans contain too many uncorrelated parameters (masses, couplings) which have to be fitted to experiment; \mbox{2) inherent} divergencies of logarithmic, linear, quadratic, $\ldots$ type need to be renormalized. The concept of {\em dynamical \mbox{generation}} \cite{Kleefeld:2008pn,Kleefeld:2005hd}
of Lagrangean theories addresses and solves both issues simultaneously:
\begin{itemize}
\item[1)] In the spirit of Eguchi \cite{Eguchi:1976iz} one starts out from very few fundamental 3-point interaction vertices and constructs then on the basis of these vertices by ``loop-shrinking''  \cite{Kleefeld:2005hd} the so-called effective action (and its underlying Lagrangean) containing also terms for all remaining n-point vertices between the fields making up the theory.
\item[2)] The couplings of the fundamental 3-point interaction vertices are then chosen such that linear, quadratic \cite{Chaichian:1995ef}, $\ldots$ divergencies cancel \cite{Kleefeld:2008pn,Kleefeld:2005hd} while the remaining logarithmic divergencies are renormalized \cite{Collins:1984xc} by adding to the effective action counter terms which replace \cite{Kleefeld:2008pn} in the spirit of the log.-divergent gap equation \cite{Delbourgo:1993dk,Delbourgo:1998kg,Hakioglu:1990kg} of Delbourgo and Scadron (DS) \cite{Delbourgo:1993dk,Delbourgo:1998kg} the integral $I_2(\overline{m}^{\,2}) \equiv \int \frac{d^4p}{(2\pi)^4} \;  \frac{1}{(p^2-\overline{m}^{\,2})^2}$ at some experimentally defined renormalization scale $\overline{m}$ (being in the case of DS approximately equal to the nonstrange constituent quark mass $\hat{m}$, i.e.\ $\overline{m}\simeq \hat{m}= m_q$) by some universal complex number $+\; \frac{i}{16\, \pi^2}$.
\end{itemize}
For various reasons like e.g. the lacking \cite{Kleefeld:2005hf,Kleefeld:2004jb,Kleefeld:2002au} evidence for the existence of gluons and new developments in mathematical physics there has developed an alternative approach to strong interactions being different from QCD which is known since the benchmarking work of DS \cite{Delbourgo:1993dk,Delbourgo:1998kg} as the Quark-Level Linear Sigma Model (QLL$\sigma$M).
\footnote{It has been the author to point out that the experimentally favoured assymptotically free phase of QLL$\sigma$M belongs to the
acceptable class of non-Hermitian PT-symmetric field theories \cite{Kleefeld:2005hd,Kleefeld:2002au,Kleefeld:2005at, Bender:1998ke}.}
Disregarding vector \& axial vector mesons the SU(2)$\times$SU(2) QLL$\sigma$M assuming $N_F=2\,N_c=6$ Fermions, one scalar isoscalar meson $\sigma$ and $N_\pi=3$ pions is constructed on the basis the interaction Lagrangean ${\cal L}_{\small\makebox{quark-meson}}(x)=g \, \overline{q^c_+}(x) \,(\sigma(x) + i \,\gamma_5 \, \vec{\tau}\cdot \vec{\pi}(x)) \,q_-(x)$ \cite{Delbourgo:1993dk,Delbourgo:1998kg,vanBeveren:2002mc} yielding by loop-shrinking the leading terms in the Lagrangean of the effective action for meson-meson interactions, i.e.\ ${\cal L}_{\small\makebox{meson-meson}} = g_{\sigma\pi\pi} \; \sigma(x) (\sigma(x)^2+\vec{\pi}(x)^2) -\frac{\lambda}{4} \, (\sigma(x)^2+\vec{\pi}(x)^2)^2 + \ldots \;$. 
Following the formalism described in Ref.\ \cite{Kleefeld:2005hd} the relevant terms in the effective action of the SU(2)$\times$SU(2) QLL$\sigma$M for the $\sigma$-one-point function (see Fig.\ \ref{fig1}), for the two-point function of the quarks (see Fig.\ \ref{fig2}), of the $\sigma$  (see Fig.\ \ref{fig3}) and of the pions  (see Fig.\ \ref{fig4}) are obtained\footnote{The result is \cite{Kleefeld:2008pn,Kleefeld:2005hd} (NLTs = Non-Local Terms): 
\begin{eqnarray}  \lefteqn{S_{(1)}[\sigma] \; = \; \int d^4x \; \sigma(x)\; i\,\Big\{  - \, 4\,g \,N_F\,m_q \; I_1(m^2_q) + 3\,g_{\sigma\pi\pi}\;I_1(m^2_\sigma)+N_\pi \,g_{\sigma\pi\pi}\;I_1(m^2_\pi)} \nonumber \\
 & + &  2\,\lambda\,g_{\sigma\pi\pi} \; i\,\Big(3\;I^{\,sunset}_{1,1,1}(m^2_\sigma,m^2_\sigma,m^2_\sigma)+N_\pi\;I^{\,sunset}_{1,1,1}(m^2_\sigma,m^2_\pi,m^2_\pi)\Big)\Big\} +  \mbox{NLTs} \; , \\[1mm]
 \lefteqn{S_{(2)}[\bar{q}q] \; = \; \frac{i}{2}\int d^4x \;\, \overline{q^c_+}(x)\,q_-(x)\; \frac{2\,g}{m^2_\sigma} \;\Big\{  - \, 4\,g \,N_F\,m_q \; I_1(m^2_q) + 3\,g_{\sigma\pi\pi}\;I_1(m^2_\sigma) + N_\pi \,g_{\sigma\pi\pi}\;I_1(m^2_\pi)} \nonumber \\
 & + & 2\,\lambda\,g_{\sigma\pi\pi} \; i\,\Big(3\;I^{\,sunset}_{1,1,1}(m^2_\sigma,m^2_\sigma,m^2_\sigma)+N_\pi\;I^{\,sunset}_{1,1,1}(m^2_\sigma,m^2_\pi,m^2_\pi)\Big)\Big\} \nonumber \\[1mm]
 & - & \frac{i}{2}\int d^4x \;\,\overline{q^c_+}(x)\;q_-(x)  \; 2\,g^2\;m_q \;\Big(  I_{1,1}(m^2_q,m^2_\sigma)- \,N_\pi\; I_{1,1}(m^2_q,m^2_\pi) \Big)+ \mbox{NLTs} \;  , \\[1mm]
 \lefteqn{S_{(3)}[\sigma^2]\; = \; \frac{i}{2}\int d^4x \; \sigma(x)^2 \;\, \frac{6\,g_{\sigma\pi\pi}}{m^2_\sigma}\, \Big\{   - \, 4\,g \,N_F\,m_q \; I_1(m^2_q) + 3\,g_{\sigma\pi\pi}\;I_1(m^2_\sigma) +N_\pi \,g_{\sigma\pi\pi}\;I_1(m^2_\pi) } \nonumber \\
 & + &  2\,\lambda\,g_{\sigma\pi\pi} \; i\,\Big(3\;I^{\,sunset}_{1,1,1}(m^2_\sigma,m^2_\sigma,m^2_\sigma)+N_\pi\;I^{\,sunset}_{1,1,1}(m^2_\sigma,m^2_\pi,m^2_\pi)\Big)\Big\} \nonumber \\
 & + & \frac{i}{2}\int d^4x \;\sigma(x)^2\;4 \,g^2 \,N_F  \; \Big( I_1(m^2_q) + 2\, m^2_q \;I_2(m^2_q) \Big) -  \frac{i}{2}\int d^4x \;\sigma(x)^2 \;\, \lambda \;\Big( 3\; I_1(m^2_\sigma) + N_\pi\; I_1(m^2_\pi) \Big)\nonumber \\
 & - & \frac{i}{2}\int d^4x \; \sigma(x)^2 \; 2\,\lambda^2\; i\;\Big(3\;I^{\,sunset}_{1,1,1}(m^2_\sigma,m^2_\sigma,m^2_\sigma)+N_\pi\;I^{\,sunset}_{1,1,1}(m^2_\sigma,m^2_\pi,m^2_\pi) \Big) \nonumber \\[1mm]
 & - & \frac{i}{2}\int d^4x  \; \sigma(x)^2 \; \,2\,g^2_{\sigma\pi\pi}\;\Big(9\;I_2(m^2_\sigma)+N_\pi\;I_2(m^2_\pi) \Big) +   \mbox{NLTs}\; , \\[1mm]
 \lefteqn{S_{(4)}[\vec{\pi}^2] \; = \; \frac{i}{2}\int d^4x \; \vec{\pi}(x)^2 \;\, \frac{2\,g_{\sigma\pi\pi}}{m^2_\sigma}\, \Big\{  - \, 4\,g \,N_F\,m_q \; I_1(m^2_q) + 3\,g_{\sigma\pi\pi}\;I_1(m^2_\sigma)} \nonumber \\
 & + & N_\pi \,g_{\sigma\pi\pi}\;I_1(m^2_\pi) + \; 2\,\lambda\,g_{\sigma\pi\pi} \; i\,\Big(3\;I^{\,sunset}_{1,1,1}(m^2_\sigma,m^2_\sigma,m^2_\sigma)+N_\pi\;I^{\,sunset}_{1,1,1}(m^2_\sigma,m^2_\pi,m^2_\pi)\Big) \Big\} \nonumber \\
 & + & \frac{i}{2}\int d^4x \;\vec{\pi}(x)^2\;\,4 \,g^2 \,N_F  \; I_1(m^2_q) - \, \frac{i}{2}\int d^4x \;\vec{\pi}(x)^2 \;\, \lambda \; \Big( I_1(m^2_\sigma) + (N_\pi+2)\; I_1(m^2_\pi) \Big)\nonumber \\[1mm]
 & - & \frac{i}{2}\int d^4x \; \vec{\pi}(x)^2 \; 2\,\lambda^2\; i\;\Big(I^{\,sunset}_{1,1,1}(m^2_\pi,m^2_\sigma,m^2_\sigma)+(N_\pi+2)\;I^{\,sunset}_{1,1,1}(m^2_\pi,m^2_\pi,m^2_\pi)\Big)\nonumber \\[1mm]
 & - & \frac{i}{2}\int d^4x  \; \vec{\pi}(x)^2 \; \,4\,g^2_{\sigma\pi\pi} \; I_{1,1}(m^2_\sigma,m^2_\pi)  +  \mbox{NLTs}\; . 
\end{eqnarray}
For convenience we have used the following short-hand notation for various relevant integrals \cite{Kleefeld:2008pn}:
\begin{eqnarray} I_n(m^2) &\equiv & \int \frac{d^4p}{(2\pi)^4} \; \frac{1}{(p^2-m^2)^n} \; \stackrel{n\ge 3}{=}\; (-1)^n\, \frac{i}{16\,\pi^2}\; \frac{1}{(n-1)!\; m^{2n-4}} \;  , \\
 I_{n_1,n_2}(m^2_1,m^2_2) & \equiv & \int \frac{d^4p}{(2\,\pi)^4}\, \frac{1}{(p^2-m^2_1)^{n_1}(p^2-m^2_2)^{n_2}} \; ,  \\[2mm]
 I^{\,sunset}_{n_1,n_2,n_3}(m_1^2,m_2^2,m^2_3)  & \equiv & \int \frac{d^4p_1}{(2\pi)^4} \;\int \frac{d^4p_2}{(2\pi)^4} \;\int \frac{d^4p_3}{(2\pi)^4} \; \; \frac{(2\pi)^4\; \delta^4 (p_1+p_2+p_3)}{(p_1^2-m_1^2)^{n_1}(p_2^2-m_2^2)^{n_2}(p_3^2-m_3^2)^{n_3}}\; .
\end{eqnarray}
In all integrals we assume the imaginary part of the squared masses to be negative.}.
After isolating the quadratically divergent part of the effective actions
it is straight forward to extract the following two conditions which yield a complete cancellation of quadratic divergencies \cite{Kleefeld:2008pn}
:\\ $0 = -4\,g\,N_F\,m_q +\left(1-\frac{\lambda}{4\pi^2}\right)g_{\sigma\pi\pi} (3+N_\pi)$ and $0 =  + 4\,g^2\,N_F -\left(1-\frac{\lambda}{4\pi^2}\right) \lambda\, (3+N_\pi)$.\\ These conditions can be finally solved for $\lambda$ and $g_{\sigma\pi\pi}$ as a function of $g$ and $m_q$ \cite{Kleefeld:2008pn}:
\begin{equation} \lambda = 2\pi^2 \left( 1 \pm \sqrt{1 - \frac{4\,g^2\,N_F}{\pi^2 (3+N_\pi)}} \;\right)\; , \;  g_{\sigma\pi\pi}  = 2\pi^2 \left( 1 \pm \sqrt{1 - \frac{4\,g^2\,N_F}{\pi^2 (3+N_\pi)}} \;\right) \; \frac{m_q}{g} \; .
\end{equation}
{\bf Acknowledgments.} It has been a great pleasure and honour to collaborate with M.D.~Scadron over many years. With this manuscript we would like to deliver our very best wishes to Mike and Arlene on the occasion of Mike's 70th birthday on February 12, 2008. This work has been supported by the
FCT of the {\em Minist\'{e}rio da Ci\^{e}ncia, Tecnologia e Ensino Superior} \/of Portugal, under Grants no.\ PDCT/FP/63907/2005, POCI/FP/81913/2007 and the Doppler and Nuclear Physics Institute (Dep.~Theor.~Phys.) at the Academy of Sciences of the Czech Republic by Project no.\ LC06002.






\clearpage

\end{document}